\documentclass[traditabstract]{aa} 
\usepackage{graphicx}
\usepackage{txfonts}
\begin{document}
   \title{Water emission from the chemically rich outflow L1157}
   \authorrunning{Vasta et al.}
   \titlerunning{Water emission towards the chemical rich outflow L1157}

   \author{M. Vasta$^1$, C. Codella$^1$, A. Lorenzani$^1$, G. Santangelo$^2$, B. Nisini$^2$, T. Giannini$^2$, M. Tafalla$^3$, R. Liseau$^4$, \\E. F. van Dishoeck$^{5,6}$, and L. Kristensen$^5$}
    \institute{ 
 Osservatorio Astrofisico di Arcetri, Largo Enrico Fermi 5, I-50125 Florence, Italy 
 \\e-mail:mvasta@arcetri.astro.it
 \and 
INAF, Osservatorio Astronomico di Roma, via di Frascati 33, 00040 Monteporzio Catone, Italy
\and 
Observatorio Astronomico Nacional (IGN), Calle Alfonso XII 3, E-28014 Madrid, Spain
\and
Department of Earth and Space Sciences, Chalmers University of Technology, Onsala Space Observatory, 439 92 Onsala, Sweden
\and
Leiden Observatory, Leiden University, P.O. Box 9513, 2300 RA Leiden, The Netherlands
\and
Max-Planck-Institut f$\ddot{\rm u}$r Extraterrestrische Physik, Giessenbachstrasse 1, 85748 Garching, Germany
}
    \date{Received date; accepted date}

\abstract {In the framework of the Herschel-WISH key program, several ortho--H$_2$O and para--H$_2$O emission lines, in the frequency range from 500 to 1700 GHz, were observed with the HIFI instrument in two bow-shock regions (B2 and R) of the L1157 cloud, which hosts what is considered to be the prototypical chemically-rich outflow.

The primary aim is to analyse water emission lines as a diagnostic of the physical conditions in the blue (B2) and red-shifted (R) lobes to compare the excitation conditions.

For this purpose, we ran the non-LTE RADEX model for a plane-parallel geometry to constrain the physical parameters (T$_{\rm kin}$, N$_{\rm H_2O}$ and n$_{\rm H_2}$) of the water emission lines detected.

A total of 5 ortho- and para-H$_2$$^{16}$O plus one o$\--$H$_2$$^{18}$O transitions were observed in B2 and R with a wide range of excitation energies (27 K$\leq$$E_{\rm u}$$\leq$215 K). The H$_2$O spectra, observed in the two shocked regions, show that the H$_2$O profiles are markedly different in the two regions. In particular, at the bow-shock R, we observed broad ($\sim$30 km s$^{-1}$ with respect to the ambient velocity) red-shifted wings where lines at different excitation peak at different red-shifted velocities.
The B2 spectra are associated with a narrower velocity range ($\sim$6 km s$^{-1}$), peaking at the systemic velocity. The excitation analysis suggests, for B2, low values of column density N$_{\rm H_2O}$ $\leq$5x10$^{13}$ cm$^{-2}$, a density range of 
10$^{5}$$\leq$n$_{\rm H_2}$$\leq$10$^{7}$ cm$^{-3}$, and warm temperatures ($\geq$300 K). 
The presence of the broad red-shifted wings and multiple peaks in the spectra of the R region, prompted the modelling of two components. High velocities are associated with relatively low temperatures ($\sim$ 100 K), N$_{\rm H_2O}$$\simeq$5x10$^{12}$--5x10$^{13}$ cm$^{-2}$ and densities n$_{\rm H_2}$$\simeq$10$^{6}$--10$^{8}$ cm$^{-3}$.
Lower velocities are associated with higher excitation conditions with T$_{\rm kin}$$\geq$300 K, very dense gas (n$_{\rm H_2}$$\sim$10$^{8}$ cm$^{-3}$) and low column density (N$_{\rm H_2O}$$<$5$\times$10$^{13}$ cm$^{-2}$). 

The overall analysis suggests that the emission in B2 comes from an extended ($\geq$15$\arcsec$) region, whilst we cannot rule out the possibility that the emission in R arises from a smaller ($>$3$\arcsec$) region. 
In this context, H$_2$O seems to be important in tracing different gas components with respect to other molecules, e.g. such as a classical jet tracer like SiO.
We have compared a grid of C- and J-type shocks spanning different velocities (10 to 40 km s$^{-1}$) and two pre-shock densities (2$\times$10$^{4}$ and 2$\times$10$^{5}$ cm$^{-3}$), with the observed intensities. Although none of these models seem to be able to reproduce the absolute intensities of the water emissions observed, it appears that the occurrence of J-shocks, which can compress the gas to very high densities, cannot be ruled out in these environments.}
 
 \maketitle
%
\section{Introduction}
Water controls the chemistry of many other species, whether in gas or solid phase and it is recognised to be 
a unique diagnostic of warm gas and energetic processes occurring in region of star formation (e.g., van Dishoeck et al. \cite{Dish}). H$_2$O is a powerful probe of the physical variations and the temporal evolution of outflowing material, which is processed by shocks produced by fast protostellar jets impacting against high-density material.
Water, unlike most molecules, cannot routinely be observed with ground-based facilities. However, space instruments, such as SWAS and Odin, have provided the first detection of the ground state line ortho-H$_2$O line at 557 GHz (e.g., Franklin et al. \cite{franco}, Bjerkeli et al. \cite{bj}). The profile of this ground state line has been compared to that of CO line. This comparison is a direct testament to the H$_2$O abundance being enhanced by many orders of magnitude in the shock (Kristensen et al. \cite{cristen}). 
ISO was also capable of detecting a number of water transitions excited in gas warmer than $\sim$80 K (e.g., Liseau et al. \cite{lise}), but with spectrally unresolved data.

The L1157 bipolar outflow is the archetypical of the so-called chemically rich outflows (Bachiller \& P\'erez Guti\'errez \cite{bp97}, hereafter BP97, Bachiller et al. \cite{bach01}). At a distance of 250 pc, and being favorably oriented in the plane of the sky, the L1157 outflow is an ideal laboratory for observations of shocks chemistry.
This outflow is known to be driven by a low-luminosity ($\sim$ 4 $L_{\odot}$) Class 0 protostar and 
it is associated with several blue-shifted (B0, B1, B2) and red-shifted 
(R0, R1, R, R2) lobes seen in CO (Gueth et al. \cite{fred98}), and in IR H$_2$ images 
(e.g. Neufeld et al. \cite{neufeld}, Nisini et al. \cite{nisi}). 

\begin{figure}[ht!]
\includegraphics[angle=-90,width=7cm]{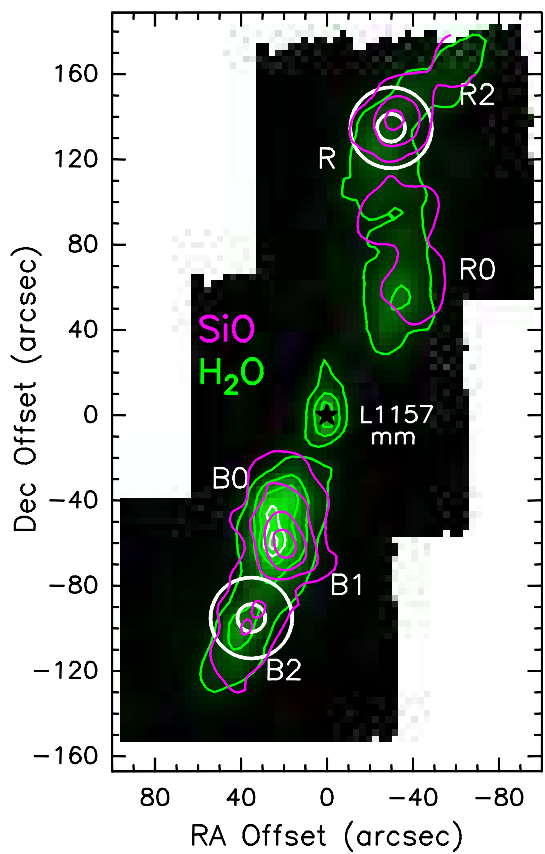}\\
\caption{PACS image of L1157 of the integrated H$_2$O emission at 1669 GHz (from Nisini et al. \cite{nisi}).
Offsets are with respect to the L1157-mm source, at coordinates $\alpha_{\rm J2000}$ = 20$^{\rm h}$ 39$^{\rm m}$ 06$\fs$2, $\delta_{\rm J2000}$ = +68$\degr$ 02$\arcmin$ 16$\farcs$0. The contours of the SiO 3-2 integrated intensity is superimposed to the water map with a spatial resolution of 18$\arcsec$ (Bachiller et al. \cite{bach01}). The positions chosen for the HIFI observations (R and B2) are marked by two white circles, which correspond to the largest and the smallest HIFI beam sizes.}
\label{mappa}
\end{figure}

The projected velocity of the northern (red-shifted) lobe is about 65$\%$ higher than that of the southern (blue-shifted) lobe (BP97) (see Fig.~\ref{mappa}).
However, the red-shifted lobe is more extended than the blue-shifted one and the kinematical ages 
of the two lobes are found to be the same (15000 yr) indicating that both lobes were created simultaneously.
Each outflow lobe shows a clumpy 
structure and it seems that for each blue clump there is a corresponding red clump symmetrically 
located with respect to the central source.
The symmetry in the individual clumps suggests that the outflow has undergone periods of enhanced 
ejection and the curved shape suggests that there have been variations in the direction of the driving wind.
In addition, the morphology is clearly S-shaped indicating the presence of an underlying precessing jet.
The brightest blue-shifted
bow-shock, B1, has been imaged at high-angular resolution revealing a 
clumpy arch-shape structure located at the top of an opened cavity (Tafalla \& Bachiller
\cite{TF95}; Gueth et al. \cite{fred98}; Benedettini et al. \cite{milena07}; Codella et al. \cite{code09}). 
L1157-B1 is well traced by species released by dust mantles such as
H$_2$CO, CH$_3$OH, and NH$_3$, complex molecules (Arce et al. \cite{arce}), and by SiO,
which is the standard tracer of high-speed shocks (e.g., Gusdorf et al. \cite{gus08b}).
In addition, L1157-B1 has been observed with HIFI and PACS spectrometers onboard the
Herschel Space Observatory (Codella et al. \cite{code10}, Lefloch et al. \cite{letter2}): the preliminary results
confirm the rich chemistry associated with the B1 position, also showing bright H$_2$O emission. 

The other bow-shocks have been less studied. As a consequence, it is still
unclear how the physical and chemical properties vary along the axis of the L1157 outflow as
well as between the blue- and red-shifted lobes.

We present the analysis of various water emission lines observed with HIFI towards two shocked regions in the L1157 outflow, which is observed as part of the the key program WISH (Water In Star-forming-regions with 
Herschel\footnote{http://www.strw.leidenuniv.nl/WISH/.}, van Dishoeck et al. \cite{Dish}).
The main aim of this paper is to constrain the physical parameters of the
gas traced by water lines at the selected B2 and R positions of L1157.
We investigate the different excitation conditions of H$_2$O in this shocked ambient gas in order to compare these conditions with
other shock tracers. Since the abundance of water can be strongly affected by shocks, we may thus infer which type of shocks exist in the outflow regions traced by water emission.

\section{Observations}
Figure~\ref{mappa} presents the PACS map of the water emission at 1669 GHz (from Nisini et al. \cite{nisi}). 
The H$_2$O map (green contour) is overlaid with contours of the emission from SiO (3-2) (magenta contour) (Bachiller et al. \cite{bach01}) transitions. The water map exhibits several emission peaks corresponding to the positions of previously-known shocked knots, labelled as B0-B1-B2 for
the south east blue-shifted lobe, and R0-R-R2 for the north west
red-shifted lobe. For our water line observations we selected the shock
positions B2 and R because they have different physical and chemical characteristics as shown in many previous works (e.g. Bachiller et al. \cite{bach01}).
Observations, from 500 to 1700 GHz, were carried out between May and November 2010 with the use of the HIFI instrument
(de Graauw et al. \cite{hifi}) in dual-beam switch mode with a nod of 3$'$ using fast chopping. The HIFI receivers are double sideband with a sideband ratio close to unity. The data were processed with the ESA-supported package HIPE\footnote{HIPE is a joint development by the {\it Herschel} Science Ground Segment Consortium, consisting of ESA, the NASA {\it Herschel} Science Center, and the HIFI, PACS and SPIRE consortia.} 5.1 ({\it Herschel} Interactive Processing Environment, Ott et al. \cite{ott}) for baseline subtraction and then exported after level 2 as FITS files in the CLASS90/GILDAS\footnote{http://www.iram.fr/IRAMFR/GILDAS} format. Two polarizations, H and V, were measured simultaneously and then averaged together to improve the signal-to-noise. We checked individual exposures for bad spectra, summed exposures, fitted a low-order polynomial baseline, and integrated line intensities as appropriate. The main beam efficiency ($\eta_{\rm mb}$) depends on frequency and is calculated as described in Roelfsema et al. (\cite{Roes}), ranging from to 0.75 to 0.71 in the 535--1670 GHz range. The absolute calibration uncertainty was estimated to be $\sim$10$\%$. At a velocity resolution of 1 km s$^{-1}$, the rms noise is 2--20 mK ($T$$_{\rm A}$ scale) for frequencies less than 1113 GHz and 60 mK at 1669 GHz.

\begin{table*}
\centering
\caption{List of molecular species and transitions observed with HIFI in L1157 B2 and R. Peak velocity, intensity
(in $T$$_{A}$ scale), integrated intensity (not corrected for beam efficiency, $F_{\rm int}$), minimum and maximum velocities (V$_{min}$, V$_{max}$), and linewidth (FWHM) are reported.}
 \scriptsize{
\label{water}
\centering
\begin{tabular}{lrccccccccl}
\hline
\multicolumn{1}{c}{Transition} &
\multicolumn{1}{c}{$\nu_{\rm 0}$$^a$} &
\multicolumn{1}{c}{$E_{\rm u}$$^a$$^,$$^b$} &
\multicolumn{1}{c}{$HPBW$} &
\multicolumn{1}{c}{$T_{\rm peak}$} &
\multicolumn{1}{c}{rms$^c$} &
\multicolumn{1}{c}{$V_{\rm peak}$} &
\multicolumn{1}{c}{$V_{\rm max}$} &
\multicolumn{1}{c}{$V_{\rm min}$} &
\multicolumn{1}{c}{$FWHM$} &
\multicolumn{1}{c}{$F_{\rm int}$} \\
\multicolumn{1}{c}{} &
\multicolumn{1}{c}{(MHz)} &
\multicolumn{1}{c}{(K)} &
\multicolumn{1}{c}{($\arcsec$)}&
\multicolumn{1}{c}{(mK)} &
\multicolumn{1}{c}{(mK)} &
\multicolumn{1}{c}{(km s$^{-1}$)} &
\multicolumn{1}{c}{(km s$^{-1}$)} &
\multicolumn{1}{c}{(km s$^{-1}$)} &
\multicolumn{1}{c}{(km s$^{-1}$)} &
\multicolumn{1}{c}{(mK km s$^{-1}$)}\\
\hline
\multicolumn{11}{c}{B2 ($\alpha_{\rm J2000}$ = 20$^{\rm h}$ 39$^{\rm m}$ 12$\fs$50, $\delta_{\rm J2000}$ = +68$\degr$ 00$\arcmin$ 41$\farcs$0)} \\
o$\--$H$_2$$^{18}$O ($1_{10}$$\--$$1_{01}$)& 547676.44&27&38&10.3(2)$^d$& 3&+1.8(0.3)$^d$& +7&--4&5.4(0.6)$^d$&59(7)$^d$\\ 
o$\--$H$_2$O ($1_{10}$$\--$$1_{01}$)& 556936.01&27&38&200(13)$^e$ & 13&+2.7(1)$^e$&+12&--9&--&8360(40)\\ 
p$\--$H$_2$O ($2_{11}$$\--$$2_{02}$)&752033.23& 137&  28&270(14)$^d$& 16&+2.4(0.2)$^d$&+7&--7&4.8(0.2)$^d$&1368(55)$^d$\\
o$\--$H$_2$O ($3_{12}$$\--$$3_{03}$)&1097364.79& 215&19& 125(17)$^d$& 19&+1.6(0.2)$^d$&+13 &--3&4.4(0.4)$^d$&583(61)$^d$\\
p$\--$H$_2$O ($1_{11}$$\--$$0_{00}$)& 1113342.96& 53& 19&75(20)$^e$ & 20&+2.6(1.0)$^e$ &+12 &--9&--&4510(33)\\
o$\--$H$_2$O ($2_{12}$$\--$$1_{01}$)& 1669904.77&  80&13&47(60)$^e$ & 60&+2.6(1.0)$^e$ &+11&--11&--&4221(120)\\ 
&&&&&&&\\
\multicolumn{11}{c}{R ($\alpha_{\rm J2000}$ = 20$^{\rm h}$ 39$^{\rm m}$ 01$\fs$00, $\delta_{\rm J2000}$ = +68$\degr$ 04$\arcmin$ 31$\farcs$0)} \\
o$\--$H$_2$O ($1_{10}$$\--$$1_{01}$)& 556936.01&27&38&256(5) & 5&+19.6(1.0)&+35&--3 &--&5027(31)\\ 
p$\--$H$_2$O ($2_{11}$$\--$$2_{02}$)&752033.23& 137& 28&176(13)& 13&+7.7(1.0)&+30&+7&--&2103(60)\\
o$\--$H$_2$O ($3_{12}$$\--$$3_{03}$)&1097364.79& 215& 19& 154(22)$^e$& 22&+6.4(1.0)$^e$&+21&--2&--&1633(99)\\
p$\--$H$_2$O ($1_{11}$$\--$$0_{00}$)& 1113342.96& 53& 19&228(15)& 15&+19.6(1.0)&+27 &--5&--&2555(74)\\
o$\--$H$_2$O ($2_{12}$$\--$$1_{01}$)& 1669904.77&  80&13&--&$<$222&--&--&--&--&--\\ 
\hline
\end{tabular}
}
\begin{center}
$^a$ Frequencies and spectroscopic parameters have been extracted from the Jet
Propulsion Laboratory molecular database (Pickett et al. \cite{pickett}). 

$^b$ With respect to the ground 
H$_2$O level $1_{01}$ 
for the ortho-H$_2$O level and $0_{00}$ 
  for the para-H$_2$O level, respectively.

$^c$ At a spectral resolution of 1 km s$^{-1}$. 

$^d$Gaussian fit when the observed profile is a gaussian-like.

$^e$Measurements taken at the dip of the emission line.
\end{center}
\end{table*}

\section{Results}
A total of 17 emission lines were detected. Table~\ref{water} lists the observed H$_2$O lines, associated with a wide range of excitation energies 
(27 K$\leq$$E_{\rm u}$$\leq$215 K). In addition, 
a detection of o$\--$H$_2$$^{18}$O $1_{10}$$\--$$1_{01}$ in B2 will help us 
to constrain the optical depth.
Also, in the observed spectral bands, additional molecular transitions have been detected in 
the B2 bow-shock, such as NH$_3$ ($1_{0}$--$0_{0}$), HCO$^{+}$(6--5), CH$_3$OH--E ($11_{1,10}$--$10_{1,9}$), and 
C$^{18}$O (5--4). On the other hand,  the $^{13}$CO (10--9) line has been observed but not detected (see Table~\ref{water_appen}). 
\subsection{Water profiles}
\begin{figure*}[ht!]
\includegraphics[angle=-90,width=16cm]{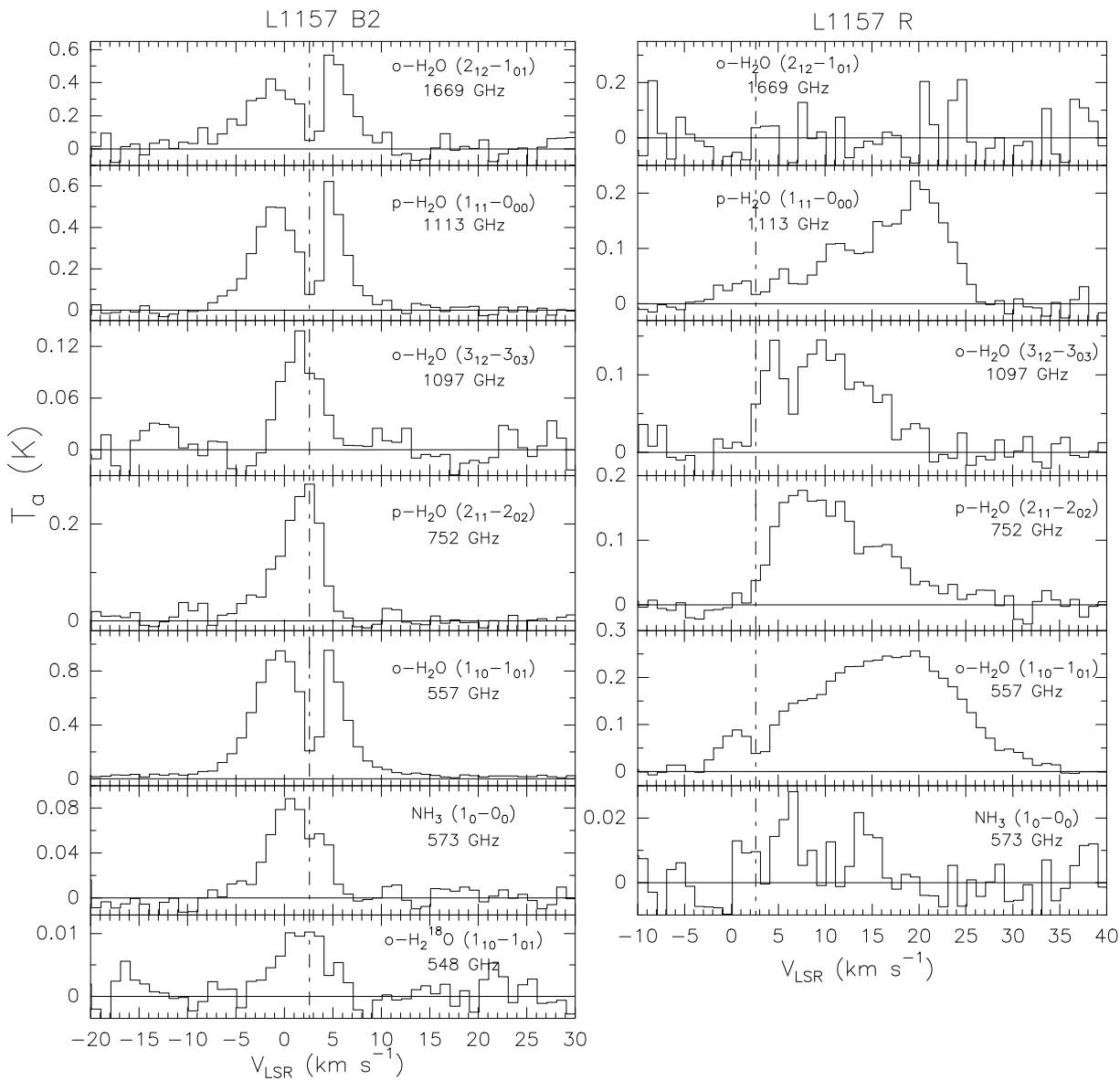}
\caption{Molecular line profiles observed towards L1157-B2 and R: species and transitions are reported in the panels. The vertical dashed line indicates the ambient LSR velocity (+ 2.6 km s$^{-1}$ from C$^{18}$O emission; BP97)}
\label{spectra}
\end{figure*}

The B2 and R water spectra shown in Fig.~\ref{spectra} (here in units of antenna temperature, $T$$_{\rm A}$) 
were smoothed to a velocity resolution of 1 km s$^{-1}$, 
to allow comparisons between all water transitions.

The spectra measured in R are consistently fainter than those measured in B2 and the water transition at 1669 GHz appears to be barely detected at R.
In the B2 spectra the profiles show a weak red emission and those lines with $E$$_{\rm u}$$\leq$80 K show absorption at ambient velocity.
The B2 spectra also show bright emission and peaks 
close to the systemic velocity while the bulk of the R emission is definitely red-shifted. 
The B2 spectra have a narrower velocity range than those observed at the R position. This could reflect a different shock velocity or simply be a consequence of a different inclination on the plane of sky. 

Interestingly, the 557 GHz profile observed at B2 is associated with a much narrower wing 
(highest blue-shifted velocity, -9 km s$^{-1}$) than that observed in B1 (-25 km s$^-1$), studied in the framework of the CHESS GT Key Project 
(see Fig. 3; Codella et al. \cite{code10}, Lefloch et al. \cite{letter2}). This could be an indication of lower shock velocities at B2. On the other hand, B2 has a broader red wing than B1, supporting the idea that the profiles are strongly affected by geometrical effects. Further comparison of H$_2$O profiles from observations obtained in B1 and B2, would be instructive and will be performed when the full set of CHESS water data will be available.

We note that the profiles in B2 narrow with increasing line excitation, while the line widths (FWZI) in R remain almost constant.
In particular, the emission produced by transitions to the ground state level are those that appear broader (see Fig.~\ref{spectra}). These spectra are also those affected by the absorption dip and whit a higher $S/N$ level. However, we cannot say if the narrow feature is a real trend.
\begin{figure}[ht!]
\includegraphics[angle=-90,width=8cm]{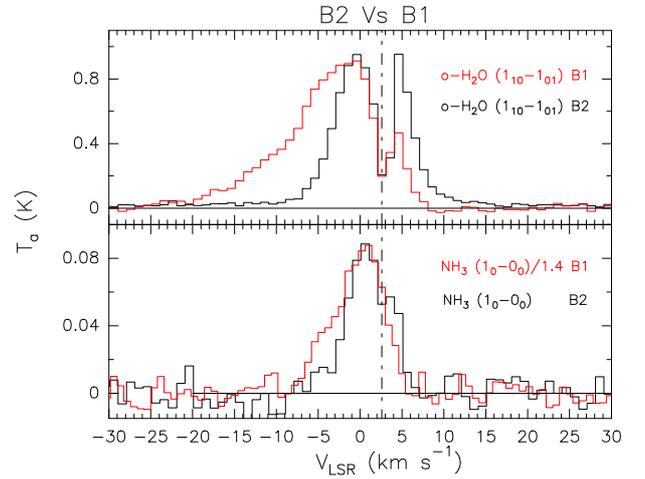}
\caption{Comparison of water emission lines at 557 GHz and NH$_3$ ($1_{0}$--$0_{0}$) observed at L1157-B1 (red profiles) and B2 (black profiles). The vertical dashed line indicates the ambient LSR velocity (+ 2.6 km s$^{-1}$ from C$^{18}$O emission; BP97)}
\label{B1vsB2}
\end{figure}

It is also notable that the spectra in R show multiple peaks that present a clear dichotomy with one peak at $\sim$+20 km s$^{-1}$ associated with transitions with $E$$_{\rm u}$$\leq$60 K and a lower velocity peak at $\sim$+10 km s$^{-1}$ associated with transitions with $E$$_{\rm u}$$\geq$136 K.
Surprisingly the high velocity emission, usually assumed to be emitted from regions close to the fast jet driving the shocks, is associated with the low excitation emission lines. 
The dichotomy observed in the R bow shock has not been detected in any of the molecules observed so far in L1157.
However, as already observed in B2, the difference in the line profiles at various excitation energies could be caused by a large amount of self-absorption in the low velocity range of those lines connecting with the ground state level. In fact, we actually observe a dip in the water emission line at 557 GHz at the systemic velocity. 
To explain such a striking difference (over 10$\--$20 km s$^{-1}$) between the line shapes, as that observed in R, the absorption would have to be caused by outflowing gas. It is evident that further multiline analysis is required to test this assumption. It should also be noted that the same dichotomy has been observed in L1148 at the R4 position by Santangelo et al. (\cite{santa}), where they clearly show that this trend is real.

\begin{figure}[ht!]
\includegraphics[angle=-90,width=8cm]{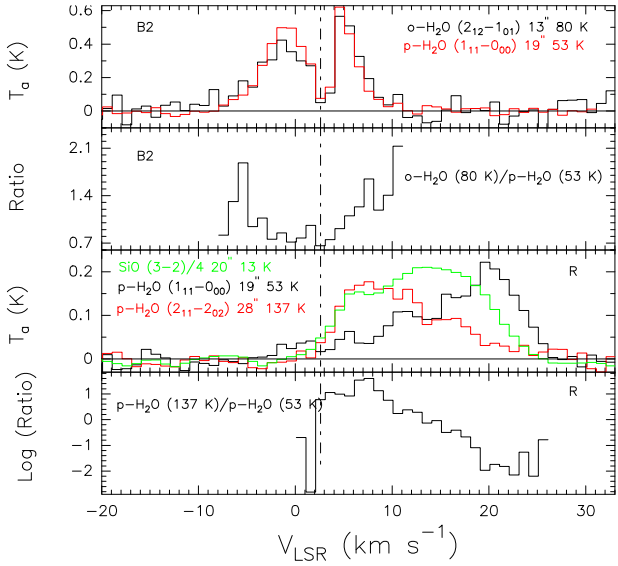}
\caption{Line temperature water ratios and water profiles in R and B2. Line temperature ratios, as function of velocity, are only shown where the S/N is larger than 3 in both emissions.}
\label{bubu}
\end{figure}

Figure~\ref{bubu} shows the intensity ratio measured towards B2 and R of transitions selected to have different excitation but observed within similar HPBW, in order to minimise effects due to beam dilution. The ratios are plotted only at those velocities at which both lines have $S/N$ $>$3. We note that, besides the obvious effects due to the presence of the absorption dip, the o$\--$H$_2$O (2$_{12}$--1$_{01}$)/p--H$_2$O (1$_{11}$--0$_{00}$) water ratio seems 
to increase with velocity, while in R it is clear that the 
p--H$_2$O (2$_{11}$--2$_{02}$)/p--H$_2$O (1$_{11}$--0$_{00}$) water ratio decreases in 
both the low and high velocity components with respect to the peak at $\sim$8 km s$^{-1}$. This effect, with an estimated line ratio error $\leq$30$\%$, reflects the distinctive dichotomy that was 
observed in the R bow-shock where water transitions with different energy excitations peak at two 
different velocities. 

In B2, the o--H$_2$O (1$_{10}$--1$_{01}$)/o--H$_2$$^{18}$O (1$_{10}$--1$_{01}$) 
ratio is used to estimate the H$_2$O optical depth, $\tau_{\rm 16}$, assuming that the two water isotopologues are
tracing the same material with the same excitation temperature. To derive the o--H$_2^{16}$O opacity, we need to use the $^{16}$O/$^{18}$O ratio for which we take the assumed local ISM value of 560 (Wilson \& Rood \cite{wilson}). The measured $\tau_{\rm 16}$ value in the low velocity line wings is about 2, which is in agreement with our non-LTE excitation analysis. However, as it cannot be confirmed that the o--H$_2^{16}$O line and the o--H$_2^{18}$O line are associated with the same excitation temperatures, any constraints placed on this value of opacity to infer the column density at B2 would be highly uncertain.

\subsection{Other line profiles}
\begin{figure}
\includegraphics[angle=-90,width=7.5cm]{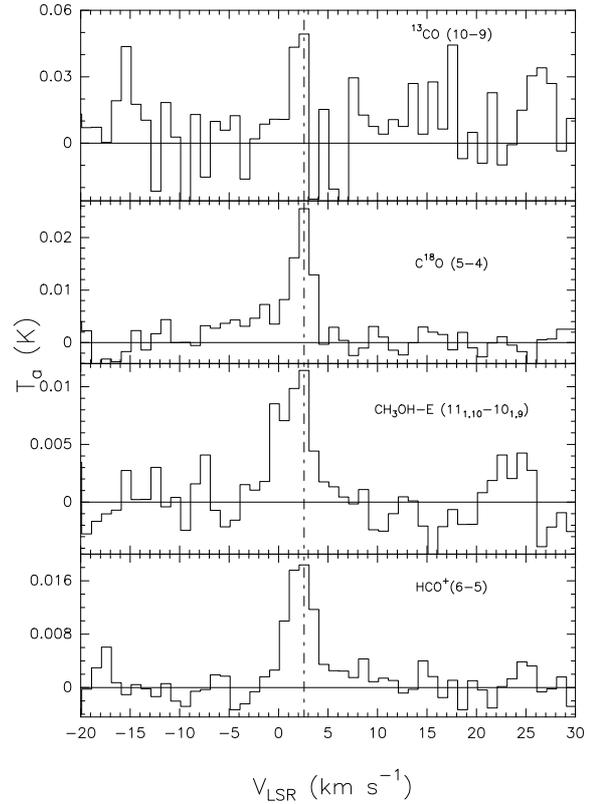}
\caption{Additional emission lines (beside the H$_2$O and NH$_3$ profiles
reported in Fig.~\ref{spectra}) observed at L1157-B2.}
\label{new}
\end{figure}

Finally, the spectral set-up allowed us to observe also the NH$_3$ ($1_{0}$--$0_{0}$) fundamental line: a tentative detection (3$\sigma$) of NH$_3$ is found at R, while relatively bright emission is detected at B2 as found in the nearby blue shock B1 
(see Fig~\ref{B1vsB2}, Codella et al. \cite{code10}). The ammonia emission, in the L1157 outflow, will be analysed in a forthcoming paper. 
Table~\ref{water_appen} lists the serendipitous detections observed at the B2 bow-shock, HCO$^{+}$ (6$\--$5), CH$_3$OH\--E ($11_{1,10}$$\--$$10_{1,9}$), $^{13}$CO (10$\--$9) and C$^{18}$O (5$\--$4), used to infer the water abundances. The B2 additional spectra are shown in Fig.~\ref{new} (here in terms of antenna temperature, $T$$_{\rm a}$). Spectra were smoothed to a velocity resolution of 1 km s$^{-1}$; they all peak at the systemic velocity and are all associated with a relatively narrow line width ($\leq$ 6 km s$^{-1}$) with respect to the water spectra.

\begin{table*}[ht!]
\centering
\caption{List of molecular species and transitions observed with HIFI in L1157 B2 and R. Peak velocity, intensity
(in $T$$_{A}$ scale), integrated intensity (not corrected for beam efficiency, $F_{\rm int}$), minimum and maximum velocities (V$_{min}$, V$_{max}$), and linewidth (FWHM) are reported.} 
 \scriptsize{
\label{water_appen}
\centering
\begin{tabular}{lrccccccccl}
\hline
\multicolumn{1}{c}{Transition} &
\multicolumn{1}{c}{$\nu_{\rm 0}$$^a$} &
\multicolumn{1}{c}{$E_{\rm u}$$^a$} &
\multicolumn{1}{c}{$HPBW$} &
\multicolumn{1}{c}{$T_{\rm peak}$} &
\multicolumn{1}{c}{rms$^b$} &
\multicolumn{1}{c}{$V_{\rm peak}$} &
\multicolumn{1}{c}{$V_{\rm max}$} &
\multicolumn{1}{c}{$V_{\rm min}$} &
\multicolumn{1}{c}{$FWHM$} &
\multicolumn{1}{c}{$F_{\rm int}$} \\
\multicolumn{1}{c}{} &
\multicolumn{1}{c}{(MHz)} &
\multicolumn{1}{c}{(K)} &
\multicolumn{1}{c}{($\arcsec$)}&
\multicolumn{1}{c}{(mK)} &
\multicolumn{1}{c}{(mK)} &
\multicolumn{1}{c}{(km s$^{-1}$)} &
\multicolumn{1}{c}{(km s$^{-1}$)} &
\multicolumn{1}{c}{(km s$^{-1}$)} &
\multicolumn{1}{c}{(km s$^{-1}$)} &
\multicolumn{1}{c}{(mK km s$^{-1}$)}\\
\hline
\multicolumn{11}{c}{Bow-shock B2} \\
HCO$^{+}$ (6$\--$5)& 535061.40&90&37&19(2)$^c$ & 3&+2.2(0.1)$^c$&+7&--2&3.3(0.3)$^c$&67(6)$^c$\\
CH$_3$OH\--E ($11_{1,10}$$\--$$10_{1,9}$)& 536191.07&169&37& 10(2)$^c$& 3&+1.6(0.3)$^c$&+7&--4&4.3(0.7)$^c$&48(6)$^c$\\
C$^{18}$O (5$\--$4)& 548830.97&79& 37&24.3(2)$^c$& 3&+2.4(0.1)$^c$&+5&--0 &2.5(0.3)$^c$&66(3)$^c$\\
NH$_3$ ($1_{0}$$\--$$0_{0}$)& 572498.07&28&38& 83.6(8)& 8&+0.9(0.2)$^b$&+7&--6 &6.2(0.4)$^b$&549(31)$^c$\\
$^{13}$CO (10$\--$9)& 1101349.65&291&21& --&20&--&--&--&--&--\\
\multicolumn{11}{c}{Bow-shock R} \\
NH$_3$ ($1_{0}$$\--$$0_{0}$)& 572498.07&28&38& 28(8)& 8&+6.6(1.0)&+9&--0  &--&105(9)\\
\hline
\end{tabular}
}
\begin{center}
$^a$ Frequencies and spectroscopic parameters have been extracted from the Jet
Propulsion Laboratory molecular database (Pickett et al. \cite{pickett}). 

$^b$ At a spectral resolution of 1 km s$^{-1}$. 

$^c$Gaussian fit when the observed profile is a gaussian-like.
\end{center}
\end{table*}

\section{Excitation Analysis}
We ran the RADEX non-LTE model (van der Tak et al. \cite{van}) with collisional rate coefficients by Faure et al. (\cite{rates})
using the escape probability method for a plane parallel geometry in order to constrain the 
physical parameters ($T$$_{\rm kin}$, $N$$_{\rm H_2O}$ and $n$$_{\rm H_2}$) of water emission. An ortho/para=3 ratio 
was assumed, equal to the high temperature equilibrium value.  
To place some constraints on the unknown size of the emitting region, we assumed three different sizes (3$\arcsec$, 15$\arcsec$ and 30$\arcsec$).
RADEX does not take into account the near-IR pumping of the H$_2$O lines where the radiation field is relatively strong and has some impact on the excitation conditions of the emitting gas. However, since both B2 and R position are relatively far from the central source, we expect no continuum to affect our results. 

\subsection{L1157 B2}
\begin{figure*}[ht!]
\includegraphics[width=9.2cm]{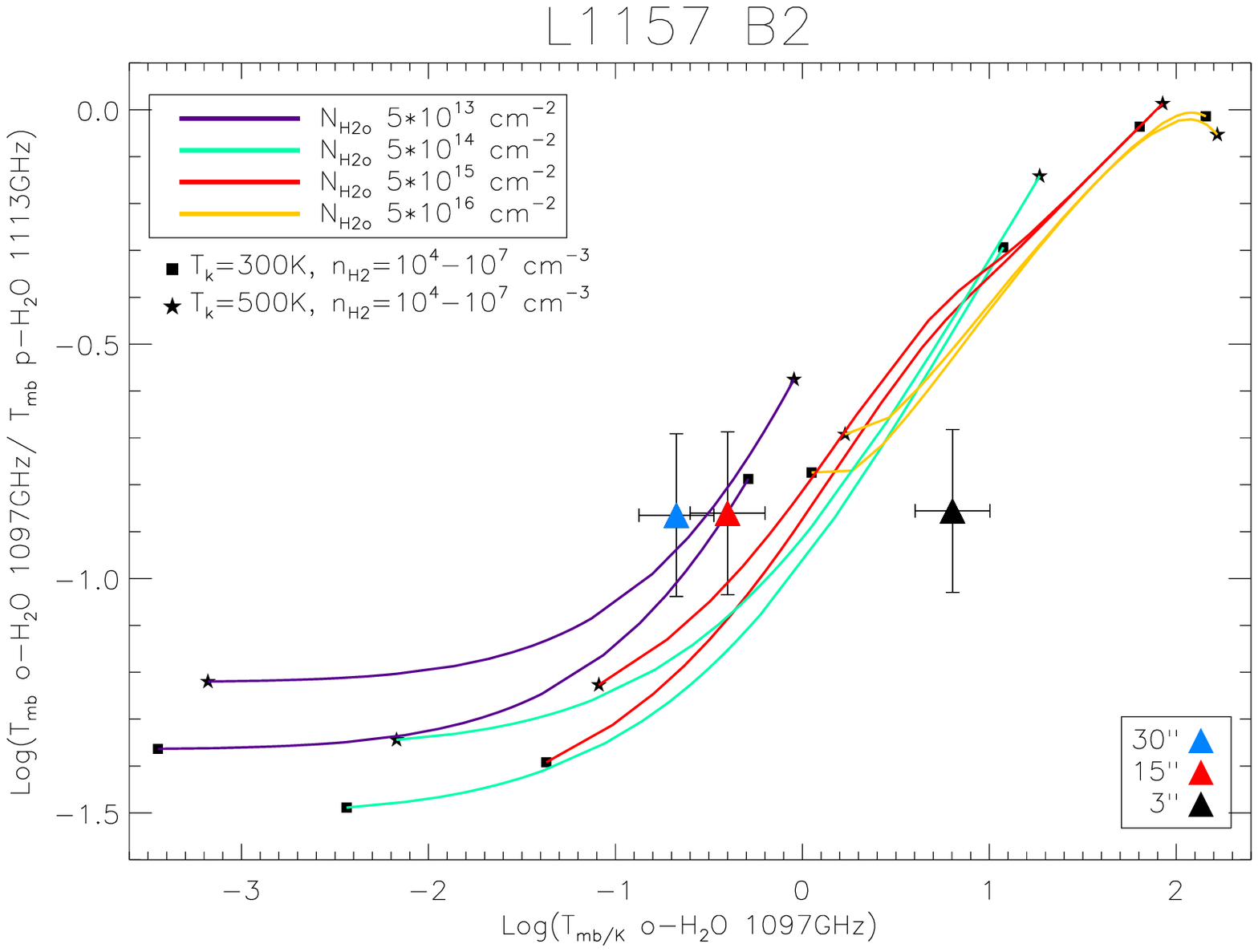}\\
\includegraphics[width=9.2cm]{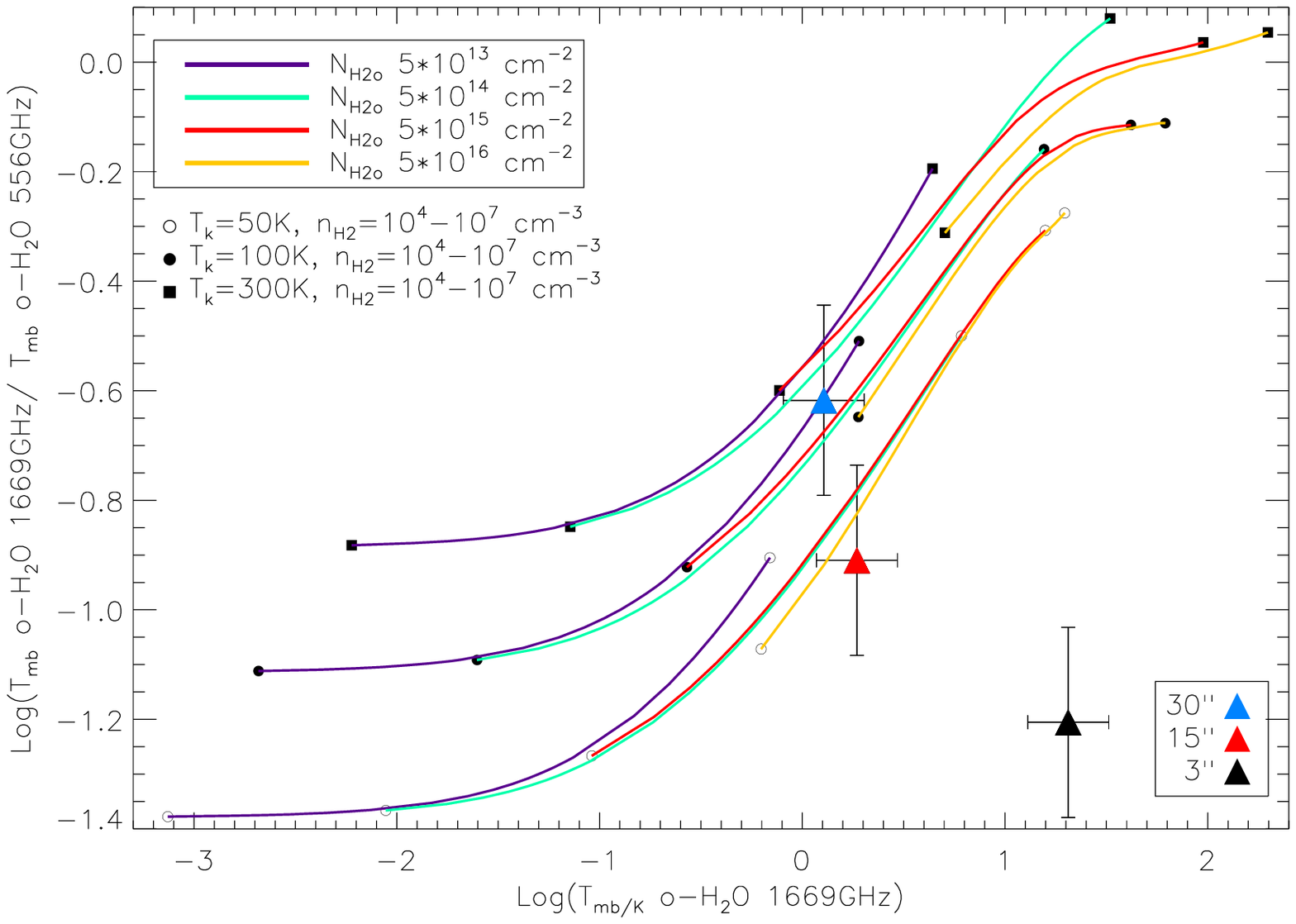}\\
\includegraphics[width=9.2cm]{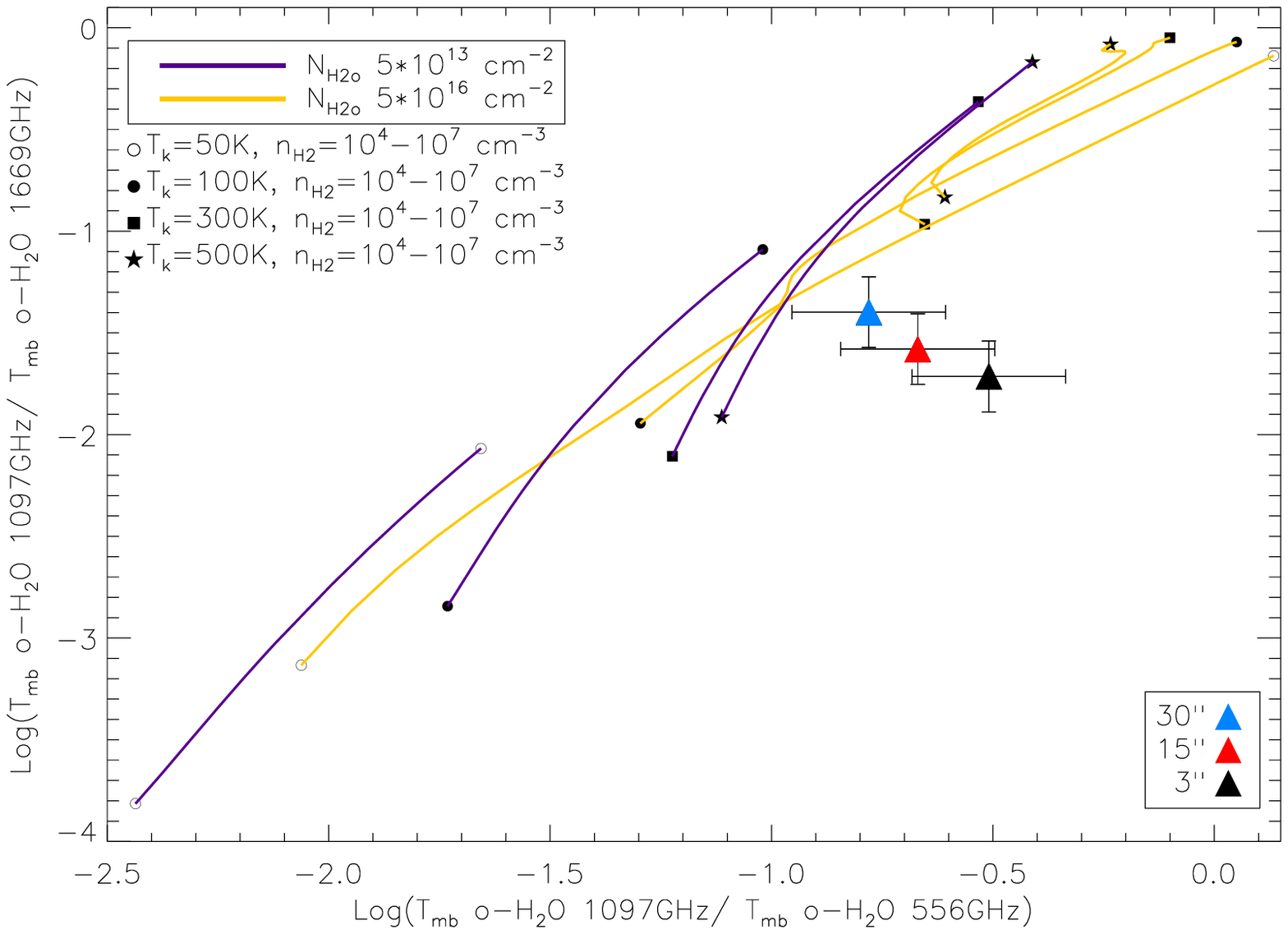}
 \caption{Analysis of the water line emissions in L1157 B2. 
o$\--$H$_2$O ($3_{12}$$\--$$3_{03}$)/p$\--$H$_2$O ($1_{11}$$\--$$0_{00}$) ratio 
versus o$\--$H$_2$O ($3_{12}$$\--$$3_{03}$) (in logarithmic scale, top panel), 
o$\--$H$_2$O ($2_{12}$$\--$$1_{01}$)/o$\--$H$_2$O ($1_{10}$$\--$$1_{01}$) ratio 
versus o$\--$H$_2$O ($2_{12}$$\--$$1_{01}$) (middle panel), 
and o$\--$H$_2$O ($3_{12}$$\--$$3_{03}$)/o$\--$H$_2$O ($1_{10}$$\--$$1_{01}$) 
versus o$\--$H$_2$O ($3_{12}$$\--$$3_{03}$)/o$\--$H$_2$O ($2_{12}$$\--$$1_{01}$) ratios (bottom panel), 
for non-LTE (RADEX) plane parallel models at the labelled temperatures. A linewidth of 5.5 km s$^{-1}$ 
has been assumed, according to the observed spectra. Each coloured curve corresponds to the 
labelled $N$$_{\rm H_2O}$ (see insert at the top-left corner). H$_2$ density increases from left to 
right as the labelled values. Triangles with error bars indicate the size of the emitting region 
(30$\arcsec$, 15$\arcsec$, 3$\arcsec$).}
\label{ortoEparaB21}
\end{figure*}

At B2, given the presence of the absorption dip and the relatively narrow velocity range, we
model the physical parameters of water emission by integrating the H$_2$O intensity over the whole spectral
profile (i.e. neglecting the absorption dip). We will consider the measurements of water transitions with an absorption dip as upper limits. 
In Fig.~\ref{ortoEparaB21} we overplot non-LTE RADEX model predictions against 
integrated flux water ratios. For this analysis, we conservatively assumed an uncertainty of 20$\%$. Coloured triangles with uncertainties 
are the observed values corrected for beam filling and assuming three different sizes as mentioned in the insert 
of the plots. 
The top panel of Fig.~\ref{ortoEparaB21} shows a plot of
the o$\--$H$_2$O ($3_{12}$$\--$$3_{03}$)/p$\--$H$_2$O ($1_{11}$$\--$$0_{00}$) ratio 
versus o$\--$H$_2$O ($3_{12}$$\--$$3_{03}$),  
the middle panel shows a plot of 
the o$\--$H$_2$O ($2_{12}$$\--$$1_{01}$)/o$\--$H$_2$O ($1_{10}$$\--$$1_{01}$) ratio 
versus o$\--$H$_2$O ($2_{12}$$\--$$1_{01}$), while the bottom panel shows a plot of 
the o$\--$H$_2$O ($3_{12}$$\--$$3_{03}$)/o$\--$H$_2$O ($2_{12}$$\--$$1_{01}$) 
versus o$\--$H$_2$O ($3_{12}$$\--$$3_{03}$)/o$\--$H$_2$O ($1_{10}$$\--$$1_{01}$) ratios.

Some general conclusions can be drawn from the inspection of Fig.~\ref{ortoEparaB21}. 
The observations are consistent with model predictions only assuming an emission size $>$3$\arcsec$ (black triangle). 
In particular, a size closer to 15$\arcsec$--30$\arcsec$ in agreement with the L1157 map 
of the o$\--$H$_2$O ($2_{12}$$\--$$1_{01}$) line published by Nisini et al. (\cite{nisi}) is 
suggested (see bottom panel).
The observed water ratio appears to be well constrained from low values of column density 
N$_{\rm H_2O}$ $\leq$5x10$^{13}$ cm$^{-2}$ (see the top panel), while a density range of 
10$^{5}$$\leq$n$_{\rm H_2}$$\leq$10$^{7}$ cm$^{-3}$, and temperatures T$_{\rm kin}$$\geq$300 K  
are inferred from all panels. 
However, because of the presence of the absorption dip in both water transitions plotted in the middle panel, the inferred physical conditions can be only considered as upper limit. In fact, the absorption dip could affect our analysis, which would shift observations towards lower excitation conditions.

With regards to the absorption dip observed in the water ground state transition, if we assume that the absorption is due to foreground
gas unrelated to L1157-B2, we estimate, from the standard radiative transfer equation, a 8 K$\leq$$T$$_{\rm ex}$$\leq$13 K adopting a 0.1$\leq$$\tau_{16}$$\leq$10. This low value of $T$$_{\rm ex}$ usually implies low densities or low kinetic temperatures. No other constraints can be inferred due to the uncertainties in the measurements and too many free parameters to fit simultaneously with the RADEX code.

\subsection{L1157 R}
\begin{figure*}[ht!]
\includegraphics[width=9.2cm]{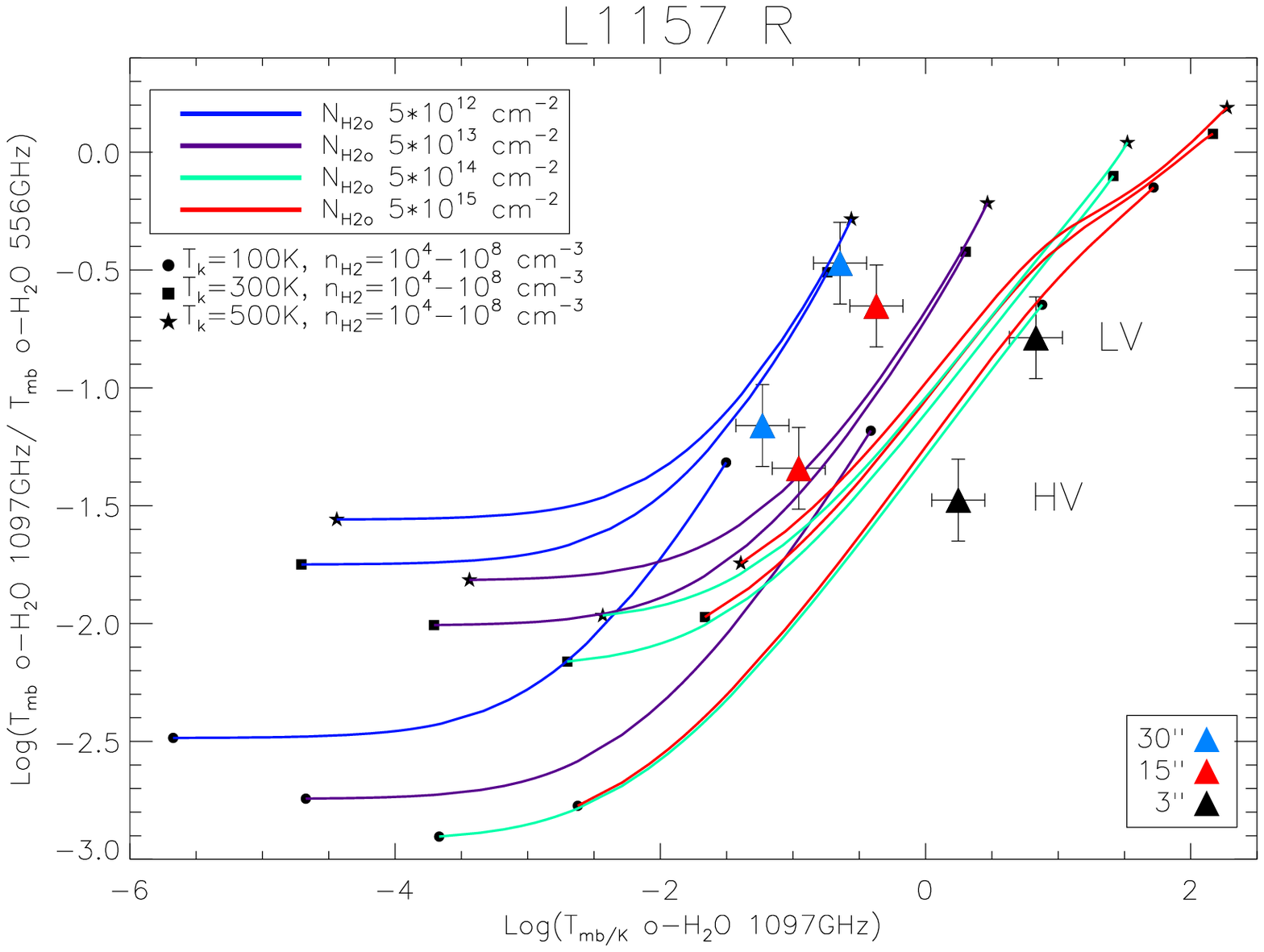}\\
\includegraphics[width=9.2cm]{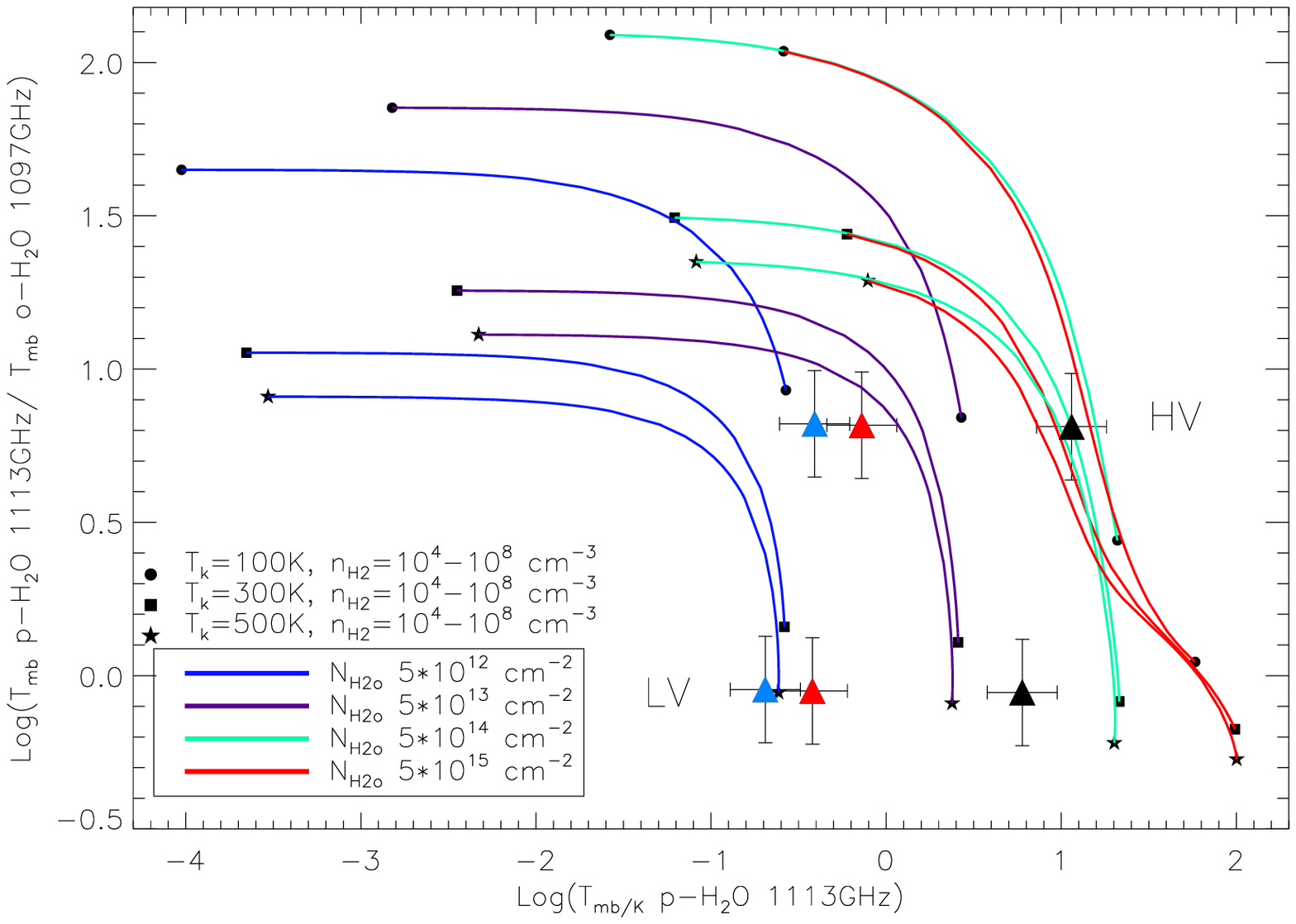}
\caption{Analysis of the water line emission in L1157 R. 
o$\--$H$_2$O ($3_{12}$$\--$$3_{03}$)/o$\--$H$_2$O ($1_{10}$$\--$$1_{01}$) ratio 
versus o$\--$H$_2$O ($3_{12}$$\--$$3_{03}$) (in logarithmic scale, top panel), 
and p$\--$H$_2$O ($1_{11}$$\--$$0_{00}$)/o$\--$H$_2$O ($3_{12}$$\--$$3_{03}$) ratio 
versus p$\--$H$_2$O ($1_{11}$$\--$$0_{00}$) (bottom panel), for non-LTE (RADEX) plane parallel  models 
at the labelled temperatures. A linewidth of 10 km s$^{-1}$ has been assumed, consistent with the 
observed spectra. Each coloured curve corresponds to the labelled $N$$_{\rm H_2O}$ 
(see insert). H$_2$ density increases 
from left to right as the labelled values. 
Triangles with error bars indicate the size of the emitting region 
(30$\arcsec$, 15$\arcsec$, 3$\arcsec$).}
\label{ortoEparaR1}
\end{figure*}

In the water profiles of the bow-shock R, Fig.~\ref{spectra} shows an evident dichotomy 
at different excitations where two emission peaks at different velocities clearly displaying different excitation
conditions.
Thus, we modelled the emission by splitting the lines into two components (hereafter called LV, at $\sim$+10 km s$^{-1}$, 
and HV, at $\sim$+20 km s$^{-1}$). 
In Fig.~\ref{ortoEparaR1} we overplotted non-LTE RADEX model 
predictions against the observed LV and HV $T$$_{\rm mb}$ water ratios. 
The top panel of Fig.~\ref{ortoEparaR1} shows 
a plot of the o$\--$H$_2$O ($3_{12}$$\--$$3_{03}$)/o$\--$H$_2$O ($1_{10}$$\--$$1_{01}$) 
ratio versus o$\--$H$_2$O ($3_{12}$$\--$$3_{03}$), while the bottom panel of Fig~\ref{ortoEparaR1} 
shows a plot of the p$\--$H$_2$O ($1_{11}$$\--$$0_{00}$)/o$\--$H$_2$O ($3_{12}$$\--$$3_{03}$) 
ratio versus p$\--$H$_2$O ($1_{11}$$\--$$0_{00}$).

A first conclusion that can be drawn is that the LV component appears to be associated with higher 
excitation conditions than the HV component and 
we can exclude a point-like source (i.e. the 3$\arcsec$ case) based on the inconsistency of column density, which can be derived from the two panels. For the low excitation HV, we have standard kinetic temperatures 
$\geq$100 K and a degeneracy between column density and hydrogen density: either 
N$_{\rm H_2O}$$\sim$5x10$^{13}$ cm$^{-2}$ and n$_{\rm H_2}$$\sim$10$^{6}$ cm$^{-3}$ or
N$_{\rm H_2O}$$\sim$5x10$^{12}$ cm$^{-2}$ and n$_{\rm H_2}$$\sim$10$^{7}$ cm$^{-3}$.
On the other hand, for the high excitation LV emission the 
observed water ratio traces temperatures T$_{\rm kin}$$\geq$300 K, as
well as surprisingly dense gas with n$_{\rm H_2}$$\sim$10$^{8}$ cm$^{-3}$
and low column density, N$_{\rm H_2O}$$<$5$\times$10$^{13}$ cm$^{-2}$. The tentative detection of NH$_3$ ($1_{0}$$\--$$0_{0}$) emission at these velocities seems to support a very high density solution.

\section{On the origin of water emission in L1157}
In summary, Table ~\ref{sum} outlines the physical conditions traced by water that we constrain from our analysis. 
\begin{table}
\caption{Physical conditions traced by water lines.} 
 \scriptsize{
\label{sum}
\begin{tabular}{lll}
\hline
B2&R-LV & R-HV \\
\hline
T$_{\rm kin}$$\geq$300 K&T$_{\rm kin}$$\geq$300 K& T$_{\rm kin}$$\geq$100 K\\
10$^{5}$$\leq$n$_{\rm H_2}$$\leq$ 10$^{7}$ cm$^{-3}$&n$_{\rm H_2}$ $\sim$ 10$^{8}$ cm$^{-3}$&N$_{\rm H_2O}$ $\sim$5x10$^{13}$ cm$^{-2}$, n$_{\rm H_2}$$\sim$10$^{6}$  cm$^{-3}$ \\
N$_{\rm H_2O}$$\leq$5x10$^{13}$ cm$^{-2}$&N$_{\rm H_2O}$$<$5x10$^{13}$ cm$^{-2}$&  N$_{\rm H_2O}$ $\sim$5x10$^{12}$ cm$^{-2}$, n$_{\rm H_2}$$\sim$10$^{7}$ cm$^{-3}$\\
Inferred size 15$\arcsec$--30$\arcsec$&Inferred size $>$3$\arcsec$&Inferred size $\geq$15$\arcsec$\\ 
\hline
\end{tabular}
}
\end{table}
The water at B2 traces lower density environments than that at R-LV and R-HV. In addition, the main difference between R-LV and R-HV is that H$_{2}$O at R-HV traces lower temperatures and densities with respect to R-LV.
The physical parameters derived in B2 and R-HV are in reasonable agreement with those measured by Nisini et al. (\cite{nisi7}), using multi transitions observations of a classical jet tracer like SiO.
On the other hand, the density inferred for R-LV (n$_{\rm H_2}$$\sim$10$^{8}$ cm$^{-3}$) is higher with respect to that derived from SiO emission by two orders of magnitude.
Actually, the SiO profiles as observed at the R position (see Fig.~\ref{bubu}) cast serious doubts on the theory that emission from these species has a common physical origin. In addition, our results strongly support the conclusion of Nisini et al. (\cite{nisi}) that water emission seems to follow quite closely the distribution of H$_2$ emission. In fact, the general physical conditions inferred in the R bow shock (T$_{\rm kin}$$\geq$300 K, n$_{\rm H_2}$$\sim$10$^{6}$--10$^{7}$ cm$^{-3}$) are consistent with those derived for the H$_2$ in Nisini et al. (\cite{nisiApj}).
\subsection{Water abundances}
We derive, at the R bow shock position, the H$_2$O abundance, with respect to H$_2$, using the $N$(H$_2$) estimated by Nisini et al. (\cite{nisi7}) and in B2 using the  C$^{18}$O (5-4) line emission detected in the present survey (see Fig.~\ref{new} and Tab.~\ref{water_appen}). 
The velocity averaged H$_2$O abundance for R is $\sim$10$^{-6}$--10$^{-7}$. 
\\The H$_2$O abundance for B2 is estimated to be $\sim$10$^{-6}$ assuming [C$^{18}$O]/[H$_{2}$]=2$\times$10$^{-7}$ (Wilson $\&$ Rood \cite{wilson}) in the temperature range from 300 to 500 K (see Table ~\ref{sum}).
This result is consistent with the water abundance found at B1 at low velocities (Lefloch et al. \cite{letter2}). On the other hand, the X(H$_2$O) measured at the highest velocities in B1 is 2 orders of magnitude greater than in B2 ($\sim$10$^{-4}$). These differences could provide evidence for an older shock in B2 compared to the one in B1.
\\The results found in R are comparable with those obtained by Santangelo et al. (\cite{santa}) in the L1448 outflow. These low water abundances could support the possibility of having a J-shock instead of a C-shock where the H$_2$O abundance is expected to be $\sim$10$^{-4}$ relative to H$_2$. Alternatively, the low water abundances measured could also be evidence of either an old shock where the water in the gas phase has had time to be depleted on grains or of UV dissociation of H$_2$O (see Bergin et al. \cite{bergi98}).
\subsection{Comparison with shocks models}
Theoretical studies indicated that, under typical interstellar conditions, shocks faster than $\sim$20 km s$^{-1}$ are efficient enough to free much of the water ice frozen on grain surfaces (Draine et al. \cite{Draine}), while shocks faster than $\sim$15 km s$^{-1}$, through gas-phase chemical reactions, produce large quantities of water (Kaufman \& Neufeld \cite{Kau}). After the shocked gas has cooled enhanced water abundances of $\sim$10$^{-4}$ relative to H$_2$ can endure for as long as 10$^{5}$ yr (Bergin et al. \cite{bergi98}). Sub-millimeter and far-infrared water emission lines are very efficient coolers, therefore large interstellar water abundances produced by the shock processing are relevant for the thermal evolution of the gas (e.g., Neufeld et al. \cite{neu95}).
Observations of a number of excited molecular transitions in L1157 clearly indicate that shocks are present in both lobes.
The present dataset deserves a more detailed comparison with shock models. In this present paper we only discuss our data using models provided by literature.

\begin{figure*}
\includegraphics[width=16cm]{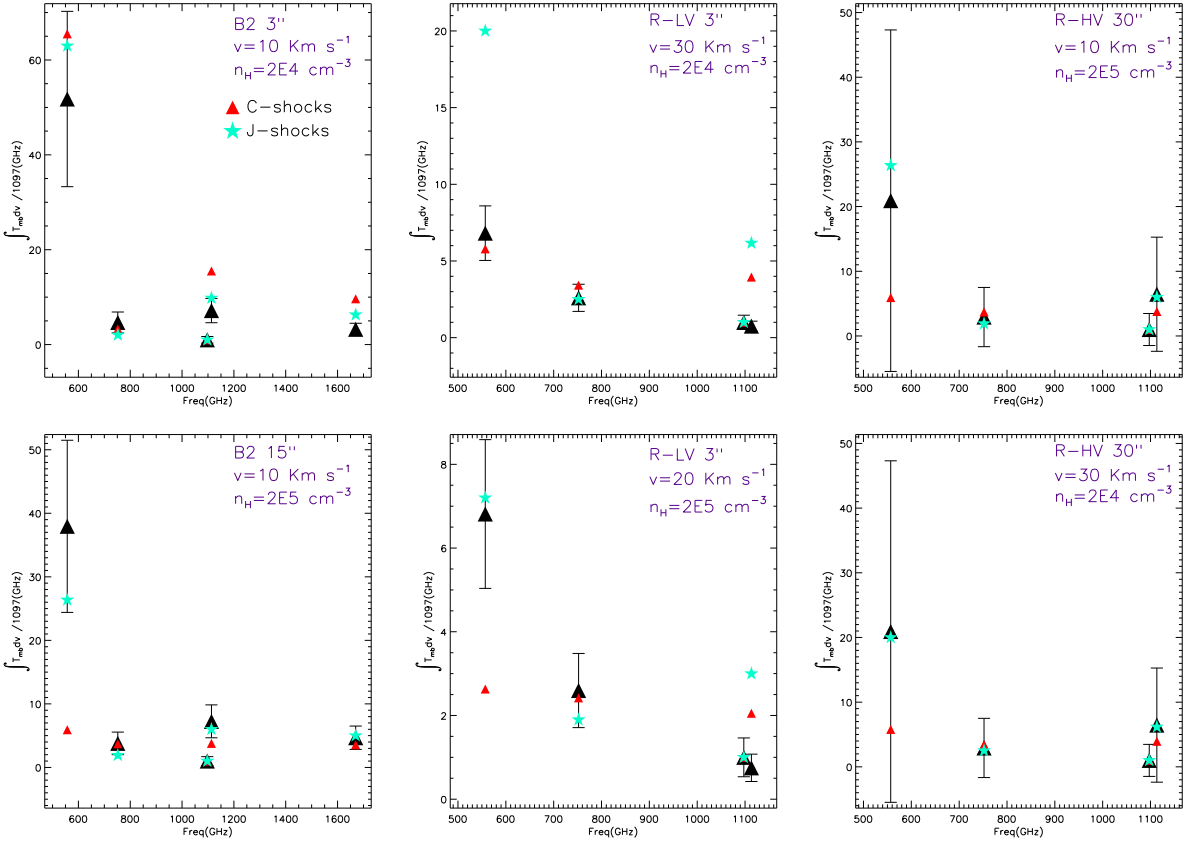}
\caption{Comparison between the observed HIFI H$_2$O line ratios of the two velocity components in R and B2 for different sizes (black triangles) and those predicted by C-type (red triangle) and J-type (cyan triangles) shock models from Flower $\&$ Pineau des For$\hat {\rm e}$ts (\cite{fiore10}).}
\label{shocks}
\end{figure*}

We have compared a grid of C- and J-type shocks spanning different velocities (10 to 40 km s$^{-1}$) and two pre-shock densities (2$\times$10$^{4}$ and 2$\times$10$^{5}$ cm$^{-3}$) provided by Flower $\&$ Pineau des For$\hat {\rm e}$ts (\cite{fiore10}), with the observed intensities.
Figure~\ref{shocks} shows the observed line ratios with respect to the high energy ($E$$_{\rm u}$=215 K) 1097 GHz line for different sizes of B2, R-LV and R-HV against the predicted shock line ratios. From our analysis, none of these models seem to be able to properly reproduce the absolute intensities of the water emissions observed.

Particularly in the case of the R-HV component, a degeneracy is noticeable because, for an emitting region of size 30$\arcsec$, both a speed of 10 km s$^{-1}$ with a pre-shock density of 2$\times$10$^{5}$ cm$^{-3}$ and a speed of 30 km s$^{-1}$ with a pre-shock density of 2$\times$10$^{4}$ cm$^{-3}$ can be inferred for both types of shock. This is consistent with the analysis made in the water diagnostic plots (Fig.~\ref{ortoEparaR1}).
The same sort of degeneracy can be drawn for the R-LV component, where we do not find a satisfactory solution that matches the full sample of water transitions.  The most plausible explanation is given by, in agreement with Fig.~\ref{ortoEparaR1}, a size of 3$\arcsec$ for a C-type shock with a speed of 30 km s$^{-1}$ and pre-shock density of 2$\times$10$^{4}$ cm$^{-3}$, and for a J-type shock with a speed of 20 km s$^{-1}$ and pre-shock density of 2$\times$10$^{5}$ cm$^{-3}$. 

Regarding the B2 position, a lower speed of J-shocks around 10 km s$^{-1}$ is inferred.
Two different solutions are found: (i) a pre-shock density of 2$\times$10$^{4}$ and 3$\arcsec$ of size, or (ii) a pre-shock density of 2$\times$10$^{5}$ and 15$\arcsec$ of size.
According to our previous water analysis an emitting size of 15$\arcsec$ looks more plausible. These preliminary results call for detailed modelling for a more complete analysis. Note, for instance, that UV H$_2$O dissociation is not taken into account in the grid of shock models used.

C-shocks have been recently invoked by Flower $\&$ Pineau des For$\hat {\rm e}$ts (\cite{fiore10}) and Viti et al. (\cite{viti}) modelling L1157 B1.
On the other hand, recent results obtained with PACS in the course of the CHESS spectral survey key-program (Benedettini et al. in prep.) required J-shocks to explain the CO excitation conditions in L1157 B1. 
This scenario could also provide a possible explanation of our analysis in B2 and R where both C- and J -type shocks seem to be required.

\section{Conclusions}
We have presented a water emission observed in the 500--1700 GHz band towards the two bow-shocks B2 and R, of the L1157 proto-stellar outflow. 
The main conclusions are the following:
\begin{enumerate}
\item
The comparison between H$_2$O and SiO profiles, and their physical characteristics casts serious doubt on the assumption that emission from these species has a similar physical origin.
\item
We derive H$_2$O abundances for R $\sim$10$^{-6}$--10$^{-7}$ and for B2 $\sim$10$^{-6}$. 
This result is consistent with the water abundance found at B1 at low velocities. On the other hand, the X(H$_2$O) measured at the highest velocities in B1 is 2 orders of magnitude greater than in B2 ($\sim$10$^{-4}$). These differences could provide evidence for an older shock in B2 compared to that in B1.
\item
The emerging scenario highlights the importance of J-shocks, expected to be associated with a thin layer and very densely compressed material in these environments. 
The results found in R are comparable with those obtained by Santangelo et al. (submitted) in the L1448 outflow. These low water abundances could support the possibility of having a J-shock instead of a C-shock.    
\item
Interestingly, the highest excitation conditions are observed at low velocities.
However, if we assume that high excitation is tracing portions of gas near the shock, we could solve this apparent contradiction by assuming that we are observing a very collimated region located along the plane of the sky. In this case the fast collimated gas should be reprojected and thus have lower radial velocities. 
On the other hand, a less excited region could be more extended, which would result in a wider velocity range due to the geometry.
\end{enumerate}

\begin{acknowledgements}
The authors are grateful to Sylvie Cabrit and the WISH internal referees
Laurent Pagani and Carolyn M$^{\textrm{c}}$Coey for their constructive comments on
the manuscript. WISH activities in Osservatorio Astrofisico di Arcetri are supported by the ASI project 01/005/11/0.
HIFI has been designed and built by a consortium of institutes and university
departments from across Europe, Canada and the United States under
the leadership of SRON Netherlands Institute for Space Research, Groningen,
The Netherlands and with major contributions from Germany, France and the
US. Consortium members are: Canada: CSA, U.Waterloo; France: CESR, LAB,
LERMA, IRAM; Germany: KOSMA, MPIfR, MPS; Ireland, NUI Maynooth;
Italy: ASI, IFSI-INAF, Osservatorio Astrofisico di Arcetri- INAF; Netherlands:
SRON, TUD; Poland: CAMK, CBK; Spain: Observatorio Astron«omico Nacional
(IGN), Centro de Astrobiologia (CSIC-INTA). Sweden: Chalmers University
of Technology - MC2, RSS \& GARD; Onsala Space Observatory; Swedish
National Space Board, Stockholm University - Stockholm Observatory;
Switzerland: ETH Zurich, FHNW; USA: Caltech, JPL, NHSC.
\end{acknowledgements}


\end{document}